\documentclass[aps,prd, reprint, showpacs,superscriptaddress,floatfix,amsmath,amssymb]{revtex4-1}

\usepackage{cancel}
\usepackage{pifont}
\usepackage{graphicx}
\usepackage{bm}
\usepackage{multirow}
\usepackage{color}
\def\apj{APJ}
\def\dSd#1{\frac{\partial S}{\partial #1}}

\begin{document}

\title{Geometrical versus wave optics under gravitational waves}

\author{Raymond Ang\'elil}\email{rangelil@physik.uzh.ch}
\affiliation{Institute for Computational Science, University of Zurich, Winterthurerstrasse 190, 8057 Zurich, Switzerland}

\author{Prasenjit Saha}
\affiliation{Physik-Institut, University of Zurich, Winterthurerstrasse 190, 8057 Zurich, Switzerland}

\date{\today}

\begin{abstract}
We present some new derivations of the effect of a plane gravitational
wave on a light ray.  A simple interpretation of the results is that a
gravitational wave causes a phase modulation of electromagnetic waves.
We arrive at this picture from two contrasting directions, namely null
geodesics and Maxwell's equations, or geometric and wave optics.
Under geometric optics, we express the geodesic equations in
Hamiltonian form and solve perturbatively for the effect of
gravitational waves.  We find that the well-known time-delay formula for
light generalizes trivially to massive particles.  We also recover, by
way of a Hamilton-Jacobi equation, the phase modulation obtained under
wave optics. Turning then to wave optics - rather than solving
Maxwell's equations directly for the fields, as in most previous
approaches - we derive a perturbed wave equation (perturbed by the
gravitational wave) for the electromagnetic four-potential.  From this
wave equation it follows that the four-potential and the electric and
magnetic fields all experience the same phase modulation.  Applying
such a phase modulation to a superposition of plane waves
corresponding to a Gaussian wave packet leads to time delays.
\end{abstract}

\maketitle

\section{Introduction}

While the existence of gravitational waves is secure from their
indirect consequences, namely, the slow orbital infall of binary
pulsars~\citep{1994RvMP...66..711T,2006Sci...314...97K}, their direct
observation is nonetheless an eagerly awaited event.

Gravitational-wave detectors seek to measure the effect of light
traveling through a ``waving'' space-time.  Interferometers with
arms that are kilometers long have been in operation for some
time~\citep{2004NIMPA.517..154A} and, with recent increases in
sensitivities~\citep{2010CQGra..27h4006H,2013arXiv1304.0670L,
  2010CQGra..27q3001A} it seems likely that detection of waves
originating from binary coalescence of stellar-mass black holes or
neutron stars may be around the corner.  Such interferometers are
commonly introduced~\citep{2005NJPh....7..204F, BhagavadGita,
  2014RvMP...86..121A} through a picture of emitter and receiver as
freely falling test particles experiencing a time-dependent
acceleration from the gravitational wave, with an interferometric
setup measuring the resulting changes in optical paths.  This simple
picture is applicable if the light path is much shorter than the
wavelength of the gravitational wave, a reasonable approximation for
ground-based interferometers.  The situation may be very different for
a space-based
interferometer~\citep{rubbo2004,2013GWN.....6....4A,2013arXiv1305.5720C,elisa},
or an experiment with spacecraft
clocks~\citep{1975GReGr...6..439E,2013CQGra..30s5011I,Giorgetta,
  earth_clox}, because these could operate in the free spectral range,
where optical path lengths are comparable to the gravitational
wavelength.  And for a pulsar timing array, consisting of a selection
of millisecond pulsars under near-continuous observation, the pulses
would have traveled through many gravitational wavelengths before
reaching the observer~\citep{FinnLommen2010}.

For a more general picture it is necessary to consider light
propagation itself through a gravitational wave space-time.  There are
two contrasting strategies for doing this, which may be described as
geometrical optics and wave optics.  In geometrical optics one
considers null geodesics in a waving space-time, and hence calculates a
light travel time.  The basic time-delay formula was first derived in the
1970s~\citep{1975GReGr...6..439E,1979ApJ...234.1100D} and has been revisited
more recently~\citep{2005PhRvD..71l2001L,2009PhRvD..80h7101C,2009CQGra..26o5010R}.
Some other early derivations are now deprecated~\citep{2009PhRvD..79b2002F} because they involve invalid shortcuts.
The transverse-traceless gauge is standard, but a gauge-invariant approach, which expresses the time delay as an integral of the Riemann
curvature along the photon path, is also possible~\citep{2014PhRvD..90f2002K}.

The use of geometrical optics, even where one wants to measure phase changes, is commonly justified as the eikonal
approximation to electromagnetic waves~\cite{2005PhRvD..71l2001L}.  Nonetheless, it is also possible
to solve Maxwell's equations in a waving space-time~\citep{1989PhLA..142..465M,1992CQGra...9.1385L,1993CQGra..10.1189C}.

This work attempts to synthesize the geometrical-optics and
wave-optics views, and to provide an intuitive description of what a gravitational wave does to an electromagnetic wave.  Such a comparison has previously been done for the small-antenna regime, that is, when the gravitational-wave phase does not change while light is traveling through the interferometer~\citep{2010AmJPh..78.1160M}.  In space-based interferometers, however, the antenna length (say $L$) may be comparable to the gravitational wavelength $2\pi/k$, and in pulsar timing $kL\gg1$.  It is thus interesting to consider the free spectral range.

\begin{figure}
\begin{center}
\includegraphics[scale = 0.55]{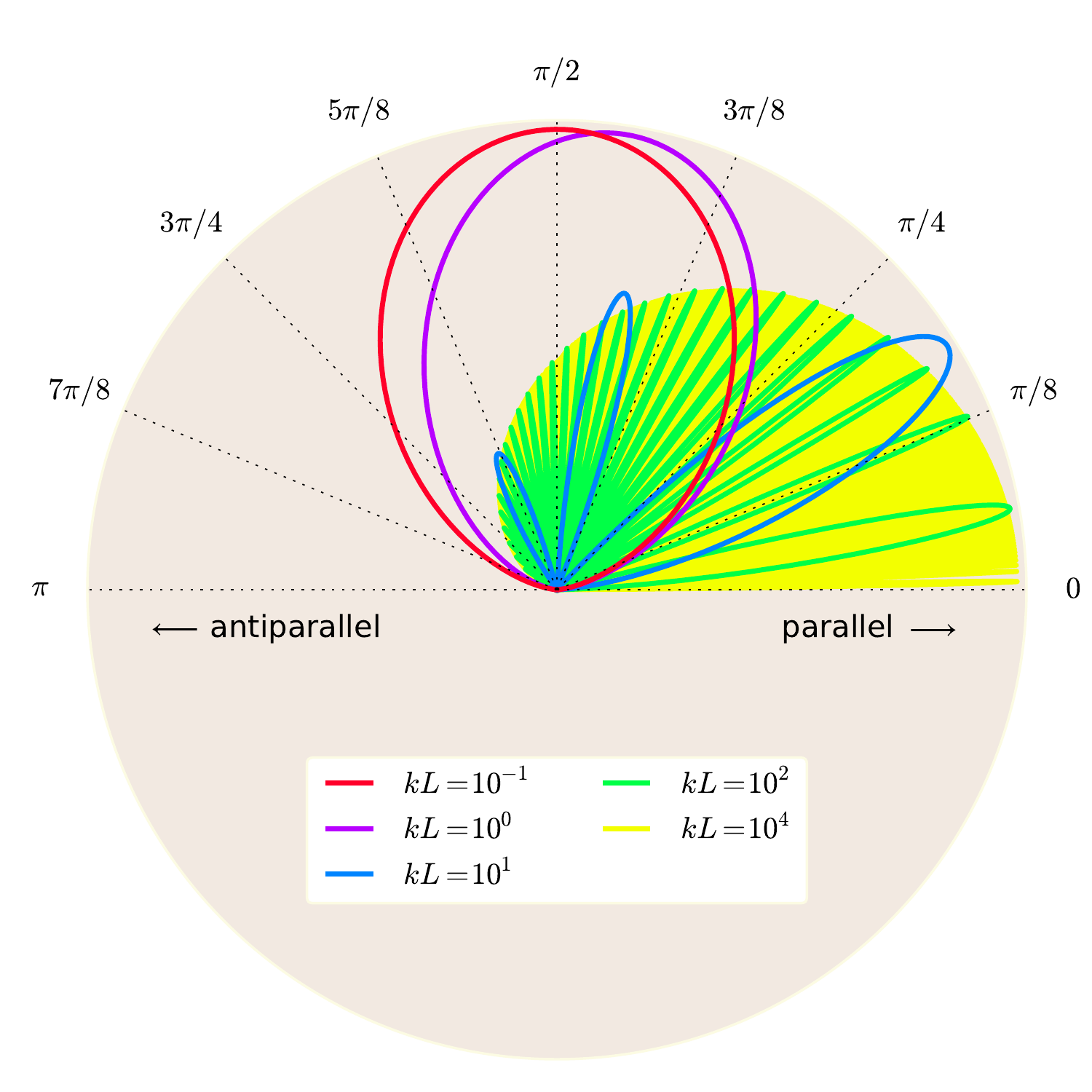}
\caption{This polar plot shows the angular dependence, $\theta$,
  between the gravitational and electromagnetic wave directions - the
  normalized form of equation \eqref{eq:phasemod}. The radial
  direction is normalized. When the detector is short compared to the
  gravitational wave period, the phase shift's directional dependence
  reduces to $\sin^2\theta$. When the detector is long (or the
  gravitational wave short), a more parallel alignment is favored, and
  an antiparallel disfavored. The maxima of the amplitudes are shown
  in Fig. 
  \ref{fig:angular_dependence_2}.}\label{fig:angular_dependence_1}
\end{center}
\end{figure}

In Sec.~\ref{sec:geom} we consider photons moving on null
geodesics.  A four-dimensional Hamiltonian formulation, with the
affine parameter being the independent variable and time being 
a coordinate, is well suited to this problem. We derive geodesics using two
different perturbative approaches.  In addition to the well-known time-delay formula, two interesting results emerge from these calculations: (i)~a generalization to massive particles and (ii)~the notion of a phase even in geometrical optics.

In Sec.~\ref{sec:wave} we consider the electromagnetic wave equation through a Minkowskian space-time perturbed by a gravitational
plane wave, and we write down a simple perturbed wave equation for the four-potential.  The perturbation terms, coupling the gravitational wave to the electromagnetic field, cause
a modulation of the phase velocity along the light path.

We show that the net effect of the gravitational wave on light is remarkably simple: the phase of the electromagnetic (EM) wave
gets modulated like
\begin{equation}\label{eq:phasemod}
\Phi = \epsilon \frac \omega{2k} \, \cos2\phi \, 
       (1+\cos\theta) \, \sin[k(t-r\cos\theta)],
\end{equation}
where we have chosen geometric units $c=1$, and $\theta$ and $\phi$ give the propagation and polarization
direction of the gravitational wave (GW) with respect to the EM wave (EMW).   The phase modulation can
be thought of as redistributing the energy of the EM wave, but no net
work is done at order $\epsilon$.  A light pulse sent from an emitter
to an observer can be expressed as a superposition of plane waves. 
Phase modulation alters the pulse shape, but the main effect is to
move the pulse, thus changing the arrival time.  For simplicity, the above expression considers a gravitational plane wave with a single wavelength, one polarization state, and a particular phase. A real detector has to contend with a superposition of such contributions.  Later in this paper, we will relate Eq. \eqref{eq:phasemod} to expressions in the literature on space-based interferometers and pulsar timing.

\begin{figure}
\begin{center}
\includegraphics[scale = 0.55]{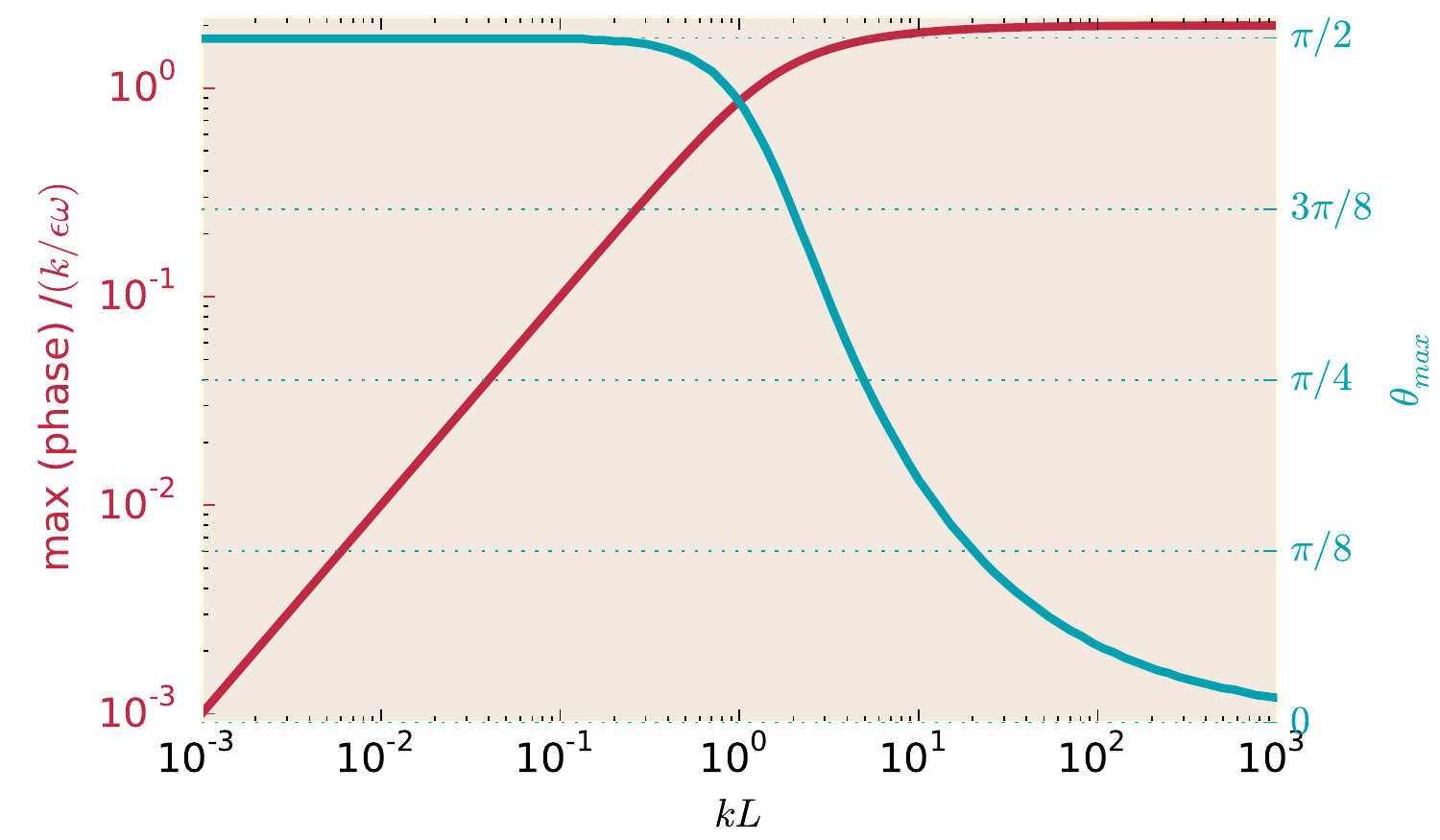}
\caption{Following from Eq. \eqref{eq:phasemod}, the red curve shows the maximum signal possible over all inclinations, $\theta$. One does not gain by increasing the detector length beyond the gravitational wave's wavelength. The blue curve shows the inclination at which this maximum occurs.}\label{fig:angular_dependence_2}
\end{center}
\end{figure}

\section{The metric} \label{sec:metric}

As in most work on gravitational waves, we will work in the transverse-traceless gauge, with the mostly pluses signature convention, in which the metric
\begin{equation}\label{mywave}
g_{\mu\nu} = \eta_{\mu\nu} +\underbrace{ \epsilon \left( \begin{array}{cccc}
0 & 0 & 0 & 0  \\
0 & 1 & 0 & 0 \\
0 & 0 & -1 & 0 \\
0 & 0 & 0 & 0 \end{array} \right) \cos\left[k\left(z - t\right)\right]}_{h_{\mu\nu}}
\end{equation}
is a weak-field solution to the vacuum gravitational field equations,
with a gravitational wave with amplitude $\epsilon$ and wave vector $k$
traveling through it.  For convenience, we have chosen the
gravitational wave to approach from the $-z$ direction, crest at the
$z=0$ plane, and bear only $\oplus$-type polarization. With these choices we retain full 
generality because at any point in the calculation we are free to
apply a spatial rotation or time shift.

For the contravariant components of the metric, one needs to
note that numerically $h^{\mu\nu}=-h_{\mu\nu}$.  For future reference,
we note the determinant of $g_{\mu\nu}$
\begin{equation}\label{gdet}
\sqrt{-g} = 1- \frac{\epsilon^2}{2}\cos^2\left(kz-kt\right)
= 1 + \mathcal{O}\left(\epsilon^2\right).
\end{equation}
We further note that the nonzero Christoffel symbols, up to
$\mathcal{O}\left(\epsilon^1 \right)$, are all the same:
\begin{equation}\label{Chsymbols}
\Gamma^x_{\;tx}= \Gamma^y_{\;ty} =\Gamma^t_{\;xx}=  \Gamma^t_{\;yy} = \frac{k\epsilon}{2}\sin\left(kz-kt\right).
\end{equation}

\section{Geometrical Optics} \label{sec:geom}

In this section we consider light to be a freely falling trajectory, i.e., one which parallel transports its tangent vector.
The gravitational wave will affect a
photon's momentum as it travels, and thereby alter the time at which the
receiver records its arrival.  A convenient formulation is in terms of
a Hamiltonian system~\cite{necronomicon}. The relevant Hamiltonian is
\begin{equation}
H = \textstyle{\frac12} p_\mu p_\nu g^{\mu\nu}
\end{equation}
or
\begin{equation} \label{hamilt}
   H = \textstyle{\frac12} \left( -p_t^2 + p_ip^i \right)
   + \textstyle{\frac12}\epsilon \left(p_y^2 - p_x^2\right) \cos(kz-kt)
\end{equation}
when we substitute the metric \eqref{mywave}.  The independent
variable is the affine parameter, say $\lambda$, and $t$ is simply a
coordinate, and $p^\mu$ is the four-momentum.  If $H=0$ the trajectory
corresponds to photons, whereas $H<0$ describes timelike geodesics.

\subsection{Null and timelike geodesics}

Since we are looking for the time delay, the Hamilton equations we need
to solve are
\begin{equation}\label{eq:crux}
\begin{aligned}
\dot{t}   &=  \frac{\partial H}{\partial p_t} = -p_t \\
\dot{p}_t &= -\frac{\partial H}{\partial t}
           = -\frac{k \epsilon}{2}\left(p_y^2 - p_x^2 \right)
             \sin\left(kz-kt\right).
\end{aligned}
\end{equation}
Let $(T,X,Y,Z,P_t,P_x,P_y,P_z)$ be a solution with no gravitational
wave.  We have
\begin{equation}\label{eq:unperturbed}
P_t  = \mbox{const,} \qquad T = - \lambda P_t
\end{equation}
and so on.  For convenience, we set $P_t=-1$, in effect choosing the
scale of $\lambda$.  Substituting the unperturbed solution in the
$O(\epsilon)$ terms in \eqref{eq:crux} and integrating, we obtain
\begin{equation}
p_t = -\frac{\epsilon}{2}\frac{ P_y^2 - P_x^2 }{P_z+P_t}
\cos\left(kZ -kT\right) - 1.
\end{equation}
Integrating once again, we arrive at
\begin{equation}
t \left(\lambda \right) = \frac{\epsilon}{2} \frac{P_y^2 - P_x^2}{P_z+P_t}
\frac{1}{k P_t}\sin\left(kZ-kZ\right) + \lambda.
\end{equation}
The time delay $\Delta t = t - T = t - \lambda$, which we may express as
\begin{equation}\label{geometric_optics_result}
\Delta t = \frac{\epsilon}{2k}\frac{p_y^2 - p_x^2}{\left(p_z + p_t\right) p_t}
\sin\left(kz-kt\right)
\end{equation}
with the understanding that the unperturbed values can be used on the
right.

Although at the beginning of this section we referred to this
trajectory as a model for light, the above derivation does not
actually require the lightlike, null stipulation $p_\mu p^\mu = 0$.
In fact, the delay is valid for null and massive trajectories alike.
Hence, the time delay \eqref{geometric_optics_result} applies also to
massive particles.  For the latter, we have
\begin{equation}
\begin{aligned}
z &= r\,\cos\theta \\
p_z &= -v p_t \, \cos\theta \\
p_x^2 - p_y^2 &= v^2 p_t^2 \, \sin^2\theta\,\sin2\phi
\end{aligned}
\end{equation}
where $v$ is the speed.  The time delay
\eqref{geometric_optics_result} can be written as
\begin{equation}\label{eq:phasemodmassive}
\frac{\epsilon}{2k} \, \frac{v^2\sin^2\theta}{1-v\cos\theta} \,
\sin\left(kr\cos\theta-kt\right).
\end{equation}
Taking the limit $v\to1$ and multiplying by the $\omega$ of the light
gives the phase change \eqref{eq:phasemod}.  For nonrelativistic
particles, the effect of gravitational waves on trajectories would be
reduced by $O(v^2)$.  It is tempting to contemplate experiments with
neutrinos shot through the Earth, with their exit times measured; but we cannot think of any plausible
scenarios for detecting gravitational waves using timelike geodesics.

Returning now to light paths, we can also rewrite the time delay in a
form similar to well-known expressions in recent literature.  Let
$\hat r^i$ be the unit vector along which a light ray travels, and let
$\hat k^i$ be the corresponding unit vector for the gravitational
wave.  Making a comparison with the metric \eqref{mywave}, we see that the time
delay \eqref{geometric_optics_result} is nothing but
\begin{equation} \label{timedelay_hij}
\Delta t = {\textstyle\frac12}
       \frac{\hat r^i \hat r^j}{1-\hat k_l \hat r^l} \int h_{ij} \, dr
\end{equation}
with the particular choice of a gravitational wave propagating in the
positive-$z$ direction in one polarization and initial phase.  The
expression may be compared with Eq.~(11) of Ref.~\cite{rubbo2004} for
space interferometers, and with Eqs.~(6) and (7) of
Ref.~\cite{FinnLommen2010} for pulsar timing.

\subsection{A Hamilton-Jacobi formulation}

An alternative perturbation method for Hamiltonian systems is to seek
a transformation from
\begin{equation}
   (x^\mu,p_\mu) \equiv (t,x^k,p_t,p_k)
\end{equation}
to a new set of canonical variables
\begin{equation}
   (X^\mu,P_\mu) \equiv (T,X^k,P_t,P_k)
\end{equation}
such that the transformed Hamiltonian (say $K$) becomes trivially
integrable. Thus
\begin{equation}
   H(p_\mu,x^\mu) = K(P_\mu) = -P_t^2 + P_k P_k \,.
\end{equation}
We allow ourselves the expression $P_kP_k$ since the transformed space
is Euclidean.  The geodesics in that space are
\begin{equation} \label{Minkowski-geod}
   T = -\lambda P_t \qquad X_k = \lambda P_k \,.
\end{equation}
As before, we take the origin as the start of the geodesics.
The canonical transformation can be specified from a generating function
$S(P_\mu,x^\mu)$, such that
\begin{equation} \label{canon-trans}
\begin{aligned}
p_\mu &= P_\mu - \dSd {x^\mu} \\
X^\mu &= x^\mu - \dSd {P_\mu}
\end{aligned}
\end{equation}
which is not one of the textbook forms, but is easy to derive, by considering
\begin{eqnarray}
p_\nu\,dx^\nu - H\,d\lambda = P_\nu\,dX^\nu &-& K\,d\lambda \nonumber  \\
- dS(P_\mu,x^\mu) &+& d(P_\nu x^\nu) - d(P_\nu X^\nu) \,.
\end{eqnarray}
The relation between the Hamiltonians is then
\begin{equation}
   K\left( p_\mu + \dSd {x^\mu} \right) = H(p_\mu,x^\mu) \,.
\end{equation}
This is a form of the Hamilton-Jacobi equation, and it has the perturbative
solution
\begin{equation} \label{basic-soln}
   K(p_\mu) + \frac{\partial K}{\partial p_\nu} \dSd {x^\nu}
   \approx H(p_\mu,x^\nu) \,.
\end{equation}

Substituting in the Hamiltonian \eqref{hamilt}, it is easy to solve for
\begin{equation} \label{eq:eikonal}
   S = \frac \epsilon{2k} \frac{p_y^2 - p_x^2}{p_z + p_t} \sin(kz-kt) \,.
\end{equation}
Since photons have $p_\mu=\hbar\omega_\mu$, this $S/\hbar$ equals the
wave-optics phase modulation \eqref{wave_optics_result}, which is the
eikonal approximation.

To calculate the time delay, we take the geodesics
\eqref{Minkowski-geod} and apply the canonical change of variable
\eqref{canon-trans}. Then 
\begin{equation} \label{tx-soln}
\begin{aligned}
   t   &= -\lambda P_t + \dSd {p_t} \\
   x^k &= \lambda P_k + \dSd {p_k}
\end{aligned}
\end{equation}
are geodesics in the original variables.  Again, we allow ourselves to
mix up and down indices here, because $P_\mu$ is Minkowskian.  Now
using the null condition
\begin{equation}
   P_t^2 = P_k P_k
\end{equation}
we have
\begin{equation}
   -\lambda P_t = \left|x^k - \dSd {p_k}\right|
\end{equation}
and with $r = |x^k|$ we can simplify the above expression to
\begin{equation}
   -\lambda P_t = r - \frac{x^k}r \dSd{p_k}
\end{equation}
to first order.  We can now write down the difference in light travel
time compared to the unperturbed case
\begin{equation}
   \Delta t \equiv t - r = \dSd {p_t} - \frac{x^k}r \dSd{p_k}
\end{equation}
which reduces to the result \eqref{geometric_optics_result} derived
from the direct method.

\section{Wave optics} \label{sec:wave}
An alternative to treating light as a geodesic path is to consider it as an electromagnetic field excitation.
We will proceed to write down Maxwell's equations in the metric
\eqref{mywave} and then find a solution for a plane electromagnetic wave
perturbed by the gravitational wave.

\subsection{Maxwell's equations}\label{curved_wave_equation}

In freely falling coordinates, Maxwell's
equations are
\begin{equation}
\nabla_\nu F^{\mu\nu} = 4\pi J^{\mu} \,
\end{equation}
where the field tensor is related to a four-potential by
\begin{equation}
F^{\mu\nu} \equiv  \partial^\mu A^\nu - \partial^\nu A^\mu.
\end{equation}
It is understood that fields and potentials are functions of $x^\mu$.
To obtain the general, global form of the equations, the partial
derivatives are promoted to covariant.  This results in
\begin{equation}\label{maxwell}
\nabla_\nu \nabla^\mu A^\nu - \nabla_\nu \nabla^\nu A^\mu = 4\pi J^{\mu}  .
\end{equation}
We can commute the covariant derivatives through the introduction of
the Riemann tensor: for an arbitrary vector field $V_\mu$,
\begin{equation}
\left[\nabla_\nu, \nabla_\gamma\right]V_{\mu} = V_\alpha R^{\alpha}_{\;\;\mu\nu\gamma}.
\end{equation}
Using this to commute the derivatives in the first term of
(\ref{maxwell}) gives
\begin{equation}\label{wupedewupe}
\underbrace{\nabla^\mu \nabla_\nu A^\nu}_{\textrm{\large\ding{202}}}  + \underbrace{R^\mu_{\;\;\alpha} A^\alpha}_{\textrm{\large\ding{203}}} - \nabla_\nu \nabla^\nu A^\mu = \underbrace{4\pi J^\mu}_{\textrm{\large\ding{204}}}.
\end{equation}
The labeled terms vanish, for the following reasons.
\begin{enumerate}
\item[\large\ding{202}] Adopting the Lorentz gauge $\nabla_\nu A^\nu = 0$ makes this  term fall away. 
\item[\large\ding{204}] We make the approximation that our
  detector's cavity is a vacuum. There are no charges in a vacuum, so
  $J^\mu = 0$;
\item[\large\ding{203}] We also assume there are no sources of stress,
  energy, or momentum, and hence the Ricci tensor components
  $R^{\mu\nu}=0$.  The electromagnetic wave itself contains some
  energy-momentum, of course, and the resulting metric perturbations
  can be expected to be of the same order as the laser power in
  dimensionless terms.  The latter quantity is $G/c^5\times[{\rm
      laser\ power}]$, and will be far smaller than even the already
  minuscule gravitational waves of interest. In other words, electromagnetic excitations do not induce significant gravitational ones.
\end{enumerate}
 
The electromagnetic field equations reduce to
\begin{equation}\label{general_PDE}
\Box_{\mathbf{g}} A^{\mu} = 0,
\end{equation}
where $\Box_\mathbf{g} \equiv \nabla_\nu \nabla^\nu$ is the covariant
d'Alembertian operator, a global generalization of $ \partial_\nu
\partial^\nu$, and can be written
\begin{equation} \label{covariant_dalembertian}
\Box_\mathbf{g} = \frac{1}{\sqrt{-g}} \frac{\partial}{\partial x^\mu} \left(\sqrt{-g} g^{\mu\nu} \partial_{\nu}   \right) .
\end{equation}
Equation (\ref{general_PDE}) are the equations of motion for the electromagnetic potentials in an arbitrary gravitational background, in the vacuum approximation. The next step is to let the gravitational wave provide the background.

\subsection{Electromagnetic wave equation in a gravitational-wave space-time}

Recalling that $\sqrt{-g}=1$ to leading order $\mathcal{O}\left(\epsilon^1\right)$ [Eq. \eqref{gdet}], the
d'Alembertian \eqref{covariant_dalembertian} reduces to
\begin{equation}
\begin{aligned}
\Box_\textbf{g} A^\lambda &= \partial_\mu \left[\left(\eta^{\mu\nu} - h^{\mu\nu} \right)\partial_\nu A^
\lambda \right] \\
&= \left[ \underbrace{\eta^{\mu\nu}\partial_\nu\partial_\mu}_{\sim 1}- \underbrace{\partial_\mu h^{\mu\nu} \partial_\nu}_{= 0} - \underbrace{h^{\mu\nu}\partial_\mu \partial_\nu }_{\sim \epsilon}\right] A^\lambda.
\end{aligned}
\end{equation}

The first term is the flat space-time wave equation. The second term is actually zero, because when $\mu, \nu = x, y$, the derivative is zero, and when $\mu, \nu = t, z$, the metric perturbation $h_{\mu\nu} = 0$. This leaves us with the third term as the only gravitational perturbation to the flat space-time result. The wave equation to solve, then, is
\begin{equation}\label{wavy-wave-equation}
\left[\Box_{\mathbf{\eta}} + \epsilon\cos\left(kz-kt\right)\left(\frac{\partial^2}{\partial x^2}-\frac{\partial^2}{\partial y^2}\right) \right] A^{\mu} = 0 .
\end{equation}
The first-term box is the Minkowskian d'Alembertian.

\subsection{Electromagnetic plane waves}

In flat space-time, an electromagnetic plane wave can be expressed as
\begin{equation}
A^\lambda = A_0^\lambda \exp \left( i\omega_\mu x_\nu \eta^{\mu\nu} \right),
\end{equation}
where the wave vector $\omega^\lambda$ satisfies the null condition
$\eta^{\mu\nu} \omega_\mu \omega_\nu = 0$.  The complex constant vector
$A_0^\lambda$ encodes the amplitude and polarization of the wave. 

It is straightforward to verify that the perturbed plane wave, that is, the real part of
\begin{equation} \label{eq:solution}
A^\lambda = A_0^\lambda \exp \left( i\omega_\mu x_\nu \eta^{\mu\nu}
                                    + i\Phi\right)
\end{equation}
where
\begin{equation}\label{wave_optics_result}
\Phi = \frac{\epsilon}{2k} \frac{\omega_x^2 - \omega_y^2}{\omega_t - \omega_z}\sin(kz-kt)
\end{equation}

satisfies the wave equation \eqref{wavy-wave-equation} to leading
order $\mathcal{O}\left(\epsilon^1\right)$.  This is an
electromagnetic plane wave with a phase modulation $\Phi$ coming from
the gravitational wave.  It is also equivalent to the geometrical
optics result \eqref{geometric_optics_result}.

To compute the components of the electric and magnetic fields, we must
evaluate the covariant derivatives.  For the electric field, we have
\begin{widetext}
\begin{equation}
\begin{aligned}
E^i &\equiv F^{ti} =
             \left[ \nabla^t A^{i} - \nabla^i A^t \right] \\
&= \left[ \partial^tA^i - \partial^i A^t +\left(\eta^{\mu t} + h^{\mu t} \right)\Gamma^i_{\;\mu\lambda} A^\lambda - \left(\eta^{\mu i}+h^{\mu i} \right)\Gamma^t_{\;\mu\lambda}A^\lambda \right] \\
&= \left[ \partial^tA^i - \partial^i A^t - \left(\Gamma^i_{\;tx}+\Gamma^t_{\;ix}\right)A^x - \left(\Gamma^i_{\;ty}+\Gamma^t_{\;iy} \right)A^y\right].
\end{aligned}
\end{equation}
\end{widetext}
In the last line, the $h^{\mu t}$ and $h^{\mu i}$ terms have been
dropped, because they are $\sim\epsilon^2$.  The nonzero Christoffel
symbols are given in \eqref{Chsymbols}.  Similarly complicated
expressions give the magnetic field.

Fortunately, there is no need to work out the covariant derivatives in
full.  One just needs to consider the different combinations of
$\omega,\epsilon,k$ that would appear in the covariant derivatives.
At zero order, there will of course be terms $\propto\omega$ as in
flat-space plane waves.  Further terms will be like:
\begin{enumerate}
\item $\mathcal{O}(\epsilon\omega^2/k)$, from the phase modulation of
  the flat-space part,
\item $\mathcal{O}(\epsilon\omega)$ from derivatives of the phase term, and
\item $\mathcal{O}(\epsilon k)$ from the Christoffel symbols.
\end{enumerate}
We assume that the light has a much shorter wavelength than the
gravitational wave (that is, $\omega\gg k$), which is surely valid for
any real or contemplated detector.  We further assume that the net
phase shift from Eq.~\eqref{wave_optics_result} is small, meaning
$\epsilon\omega/k\ll1$.  Accordingly, we have
\begin{equation}
\omega \gg \epsilon \omega^2/k \gg \epsilon\omega \gg \epsilon k.
\end{equation}
Hence, to leading order, the electric and magnetic field will receive
the same phase modulation as the four-potential. In fact, all
covariant first derivatives of the potential receive the same
modulation as the potential. This also means that our solution
respects the Lorentz gauge condition.

In other words, the phase modulation \eqref{wave_optics_result} is the
leading-order effect of the gravitational wave on the electromagnetic
field.  If we now write
\begin{equation}
\omega_x^2 - \omega_y^2 = \omega^2\sin^2\!\theta\,\cos2\phi, \quad
\omega_z = \omega\cos\theta, \quad z = r\cos\theta
\end{equation}
we obtain the expression \eqref{eq:phasemod} for the phase. The angular dependence we find is confirmed by gravitational wave antenna theory (see Maggiore (2008) cf. Eq 9.136~\citep{maggiore}).

\subsection{Pulses}
The solution \eqref{eq:solution} corresponds to a single plane wave. Just like in Minkowski space, the linearity of the electromagnetic field equations in a curved space-time \eqref{general_PDE} means we may stack solutions corresponding to whatever the light source does. We may replace \eqref{eq:solution} with its Fourier transformation - i.e. an integral over all frequency amplitudes. Consider for example, an experiment in which a beam is pulsed at regular intervals. This may be modeled as an isolated portion of light, emitted at a definite time from the emitter, which travels down the cavity to arrive at the observer and,
depending on the phase, will be observed to arrive either before or
after a Minkowski photon. If we let the pulse take on the Gaussian
wave-packet form at the emitter, in light-frequency space it is
\begin{equation}
A\left(w \right) = \exp\left[-\alpha\left(\omega-\omega_0 \right)^2\right],
\end{equation}
and in real space
\begin{equation} \label{eq:pulse}
A\left(x^\mu\right) \sim \cos\left[\omega_0 \left(r - t + \Phi \right)\right] \exp \left[ \frac{1}{4\alpha} \left(r - t + \Phi\right)^2 \right],
\end{equation}
with $\Phi$ given by \eqref{wave_optics_result}. This traveling wave packet's motion is modulated by the gravitational wave, affecting its arrival time at the observer. Because each frequency component interacts with the gravitational wave differently \eqref{eq:phasemod}, the pulse shape changes as it travels.

\section{Discussion}

\begin{figure*}
\begin{center}
\includegraphics[scale = 0.3]{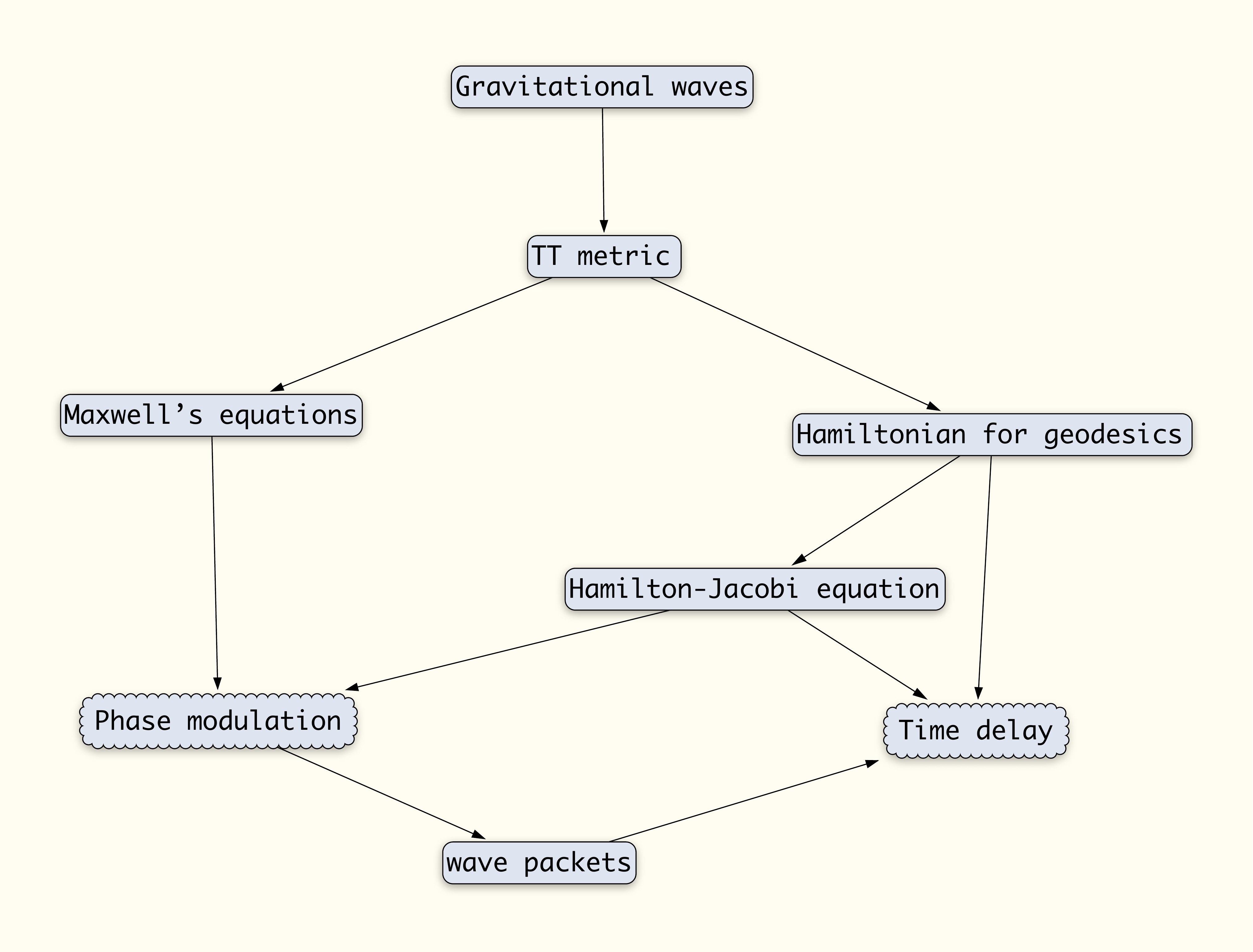}
\caption{A mind map showing the relationships between some of the concepts discussed in this paper. There are two routes each to the two observables: time delay, and phase modulation.}\label{fig:MINDMAP}
\end{center}
\end{figure*}

Figure \ref{fig:MINDMAP} summarizes, in mind-map style, the
relationship between concepts and their progression as explored in
this paper.  The essential results are Eq. \eqref{eq:phasemod} or
Eq. \eqref{eq:eikonal} for the phase modulation, and Eqs.
\eqref{geometric_optics_result} or \eqref{eq:pulse} for time delays.
The well-known schematic picture of a gravitational wave detector in
terms of a ring of freely falling test particles is avoided, and the
conundrums  associated with that
picture~\citep{1997AmJPh..65..501S,faoroni} do not appear.

As an alternative picture, we imagine listening to a bird song at a
rock concert: the bass thump produces local changes in the sound
speed, inducing a pseudo-Doppler shift in the bird song. Or, in our case, the gravitational wave modulates space-time's refractive index. 

Another correspondence is possible with Alcubierre's warp drive~\citep{warpdrive} space-time, consisting of a time-dependent metric for a kind of cavity which skims atop a Minkowski background. The respective local contraction and expansion of space-time in front of and behind the moving cavity permit the special relativity-abiding passengers of the bubble to experience arbitrarily short travel times \footnote{This metric solution to the field equations is unrealistic because it requires energy density distributions which are unphysical because they violate responsible energy conditions.}. The Alcubierre approach to faster-than-light travel is not unlike the scenario we have described in this paper. The waving space-time provides regions of time-dependent contraction and expansion. From the wave-optics perspective, the phase velocity \eqref{eq:solution} modulates about $c$. A light pulse shot down a waving space-time at just the right time can arrive at its destination earlier than if the space-time were Minkowski - i.e., if no wave were present.  An interesting outcome of the geometric optics approach \eqref{geometric_optics_result} is that the expression for the time delay holds even when the trajectory is not null, i.e. it generalizes trivially to massive bodies. In other words, gravitational waves allow for travel times very slightly shorter than what is permitted by special relativity, even for spaceships.  

The simple form of the solutions we present mean that they are suitable for illustrations as animations. In the Supplemental Material to this paper we include two user-interactive animations written in Python using the matplotlib library~\citep{matplotlib}. They animate the time evolution of the electromagnetic field as it travels down a cavity in a ``'waving space-time''. One animation represents monochromatic laser light (phase shifts), and the other a pulsed beam (time delays). 

Solving the EMW equation under GW modulation \eqref{wavy-wave-equation} under more exotic boundary conditions would be intriguing: for example examining how a waveguide's propagation modes are altered by the gravitational wave.
It would be interesting to apply the phase-modulation picture
to extend time-delay interferometry, with unequal interferometer arms,
moving sources and all~\cite{lrr-2005-4}, to the free spectral
range. It is possible that phase locking of the detector
cavity gives rise to resonances, which future instruments could
exploit. 

\begin{acknowledgments}
We thank L.S. Finn for the insightful discussion, and C\'edric Huwyler and Andreas Sch\"arer for helping clarify certain points. We thank the referee for the useful comments. R.A. acknowledges support from the Swiss National Science Foundation.
\end{acknowledgments}

\bibliographystyle{apsrev4-1}
\bibliography{heap}

\begin{thebibliography}{36}%
\makeatletter
\providecommand \@ifxundefined [1]{%
 \@ifx{#1\undefined}
}%
\providecommand \@ifnum [1]{%
 \ifnum #1\expandafter \@firstoftwo
 \else \expandafter \@secondoftwo
 \fi
}%
\providecommand \@ifx [1]{%
 \ifx #1\expandafter \@firstoftwo
 \else \expandafter \@secondoftwo
 \fi
}%
\providecommand \natexlab [1]{#1}%
\providecommand \enquote  [1]{``#1''}%
\providecommand \bibnamefont  [1]{#1}%
\providecommand \bibfnamefont [1]{#1}%
\providecommand \citenamefont [1]{#1}%
\providecommand \href@noop [0]{\@secondoftwo}%
\providecommand \href [0]{\begingroup \@sanitize@url \@href}%
\providecommand \@href[1]{\@@startlink{#1}\@@href}%
\providecommand \@@href[1]{\endgroup#1\@@endlink}%
\providecommand \@sanitize@url [0]{\catcode `\\12\catcode `\$12\catcode
  `\&12\catcode `\#12\catcode `\^12\catcode `\_12\catcode `\%12\relax}%
\providecommand \@@startlink[1]{}%
\providecommand \@@endlink[0]{}%
\providecommand \url  [0]{\begingroup\@sanitize@url \@url }%
\providecommand \@url [1]{\endgroup\@href {#1}{\urlprefix }}%
\providecommand \urlprefix  [0]{URL }%
\providecommand \Eprint [0]{\href }%
\providecommand \doibase [0]{http://dx.doi.org/}%
\providecommand \selectlanguage [0]{\@gobble}%
\providecommand \bibinfo  [0]{\@secondoftwo}%
\providecommand \bibfield  [0]{\@secondoftwo}%
\providecommand \translation [1]{[#1]}%
\providecommand \BibitemOpen [0]{}%
\providecommand \bibitemStop [0]{}%
\providecommand \bibitemNoStop [0]{.\EOS\space}%
\providecommand \EOS [0]{\spacefactor3000\relax}%
\providecommand \BibitemShut  [1]{\csname bibitem#1\endcsname}%
\let\auto@bib@innerbib\@empty
\bibitem [{\citenamefont {{Taylor}}(1994)}]{1994RvMP...66..711T}%
  \BibitemOpen
  \bibfield  {author} {\bibinfo {author} {\bibfnamefont {J.~H.}\ \bibnamefont
  {{Taylor}}, \bibfnamefont {Jr.}},\ }\href {\doibase
  10.1103/RevModPhys.66.711} {\bibfield  {journal} {\bibinfo  {journal}
  {Reviews of Modern Physics}\ }\textbf {\bibinfo {volume} {66}},\ \bibinfo
  {pages} {711} (\bibinfo {year} {1994})}\BibitemShut {NoStop}%
\bibitem [{\citenamefont {{Kramer}}\ \emph {et~al.}(2006)\citenamefont
  {{Kramer}}, \citenamefont {{Stairs}}, \citenamefont {{Manchester}},
  \citenamefont {{McLaughlin}}, \citenamefont {{Lyne}}, \citenamefont
  {{Ferdman}}, \citenamefont {{Burgay}}, \citenamefont {{Lorimer}},
  \citenamefont {{Possenti}}, \citenamefont {{D'Amico}}, \citenamefont
  {{Sarkissian}}, \citenamefont {{Hobbs}}, \citenamefont {{Reynolds}},
  \citenamefont {{Freire}},\ and\ \citenamefont
  {{Camilo}}}]{2006Sci...314...97K}%
  \BibitemOpen
  \bibfield  {author} {\bibinfo {author} {\bibfnamefont {M.}~\bibnamefont
  {{Kramer}}}, \bibinfo {author} {\bibfnamefont {I.~H.}\ \bibnamefont
  {{Stairs}}}, \bibinfo {author} {\bibfnamefont {R.~N.}\ \bibnamefont
  {{Manchester}}}, \bibinfo {author} {\bibfnamefont {M.~A.}\ \bibnamefont
  {{McLaughlin}}}, \bibinfo {author} {\bibfnamefont {A.~G.}\ \bibnamefont
  {{Lyne}}}, \bibinfo {author} {\bibfnamefont {R.~D.}\ \bibnamefont
  {{Ferdman}}}, \bibinfo {author} {\bibfnamefont {M.}~\bibnamefont {{Burgay}}},
  \bibinfo {author} {\bibfnamefont {D.~R.}\ \bibnamefont {{Lorimer}}}, \bibinfo
  {author} {\bibfnamefont {A.}~\bibnamefont {{Possenti}}}, \bibinfo {author}
  {\bibfnamefont {N.}~\bibnamefont {{D'Amico}}}, \bibinfo {author}
  {\bibfnamefont {J.~M.}\ \bibnamefont {{Sarkissian}}}, \bibinfo {author}
  {\bibfnamefont {G.~B.}\ \bibnamefont {{Hobbs}}}, \bibinfo {author}
  {\bibfnamefont {J.~E.}\ \bibnamefont {{Reynolds}}}, \bibinfo {author}
  {\bibfnamefont {P.~C.~C.}\ \bibnamefont {{Freire}}}, \ and\ \bibinfo {author}
  {\bibfnamefont {F.}~\bibnamefont {{Camilo}}},\ }\href {\doibase
  10.1126/science.1132305} {\bibfield  {journal} {\bibinfo  {journal}
  {Science}\ }\textbf {\bibinfo {volume} {314}},\ \bibinfo {pages} {97}
  (\bibinfo {year} {2006})},\ \Eprint {http://arxiv.org/abs/astro-ph/0609417}
  {astro-ph/0609417} \BibitemShut {NoStop}%
\bibitem [{\citenamefont {{Abbott}}\ \emph {et~al.}(2004)\citenamefont
  {{Abbott}}, \citenamefont {{Abbott}}, \citenamefont {{Adhikari}},
  \citenamefont {{Ageev}}, \citenamefont {{Allen}}, \citenamefont {{Amin}},
  \citenamefont {{Anderson}}, \citenamefont {{Anderson}}, \citenamefont
  {{Araya}}, \citenamefont {{Armandula}},\ and\ \citenamefont
  {et~al.}}]{2004NIMPA.517..154A}%
  \BibitemOpen
  \bibfield  {author} {\bibinfo {author} {\bibfnamefont {B.}~\bibnamefont
  {{Abbott}}}, \bibinfo {author} {\bibfnamefont {R.}~\bibnamefont {{Abbott}}},
  \bibinfo {author} {\bibfnamefont {R.}~\bibnamefont {{Adhikari}}}, \bibinfo
  {author} {\bibfnamefont {A.}~\bibnamefont {{Ageev}}}, \bibinfo {author}
  {\bibfnamefont {B.}~\bibnamefont {{Allen}}}, \bibinfo {author} {\bibfnamefont
  {R.}~\bibnamefont {{Amin}}}, \bibinfo {author} {\bibfnamefont {S.~B.}\
  \bibnamefont {{Anderson}}}, \bibinfo {author} {\bibfnamefont {W.~G.}\
  \bibnamefont {{Anderson}}}, \bibinfo {author} {\bibfnamefont
  {M.}~\bibnamefont {{Araya}}}, \bibinfo {author} {\bibfnamefont
  {H.}~\bibnamefont {{Armandula}}}, \ and\ \bibinfo {author} {\bibnamefont
  {et~al.}},\ }\href {\doibase 10.1016/j.nima.2003.11.124} {\bibfield
  {journal} {\bibinfo  {journal} {Nuclear Instruments and Methods in Physics
  Research A}\ }\textbf {\bibinfo {volume} {517}},\ \bibinfo {pages} {154}
  (\bibinfo {year} {2004})},\ \Eprint {http://arxiv.org/abs/gr-qc/0308043}
  {gr-qc/0308043} \BibitemShut {NoStop}%
\bibitem [{\citenamefont {{Harry}}\ and\ \citenamefont {{LIGO Scientific
  Collaboration}}(2010)}]{2010CQGra..27h4006H}%
  \BibitemOpen
  \bibfield  {author} {\bibinfo {author} {\bibfnamefont {G.~M.}\ \bibnamefont
  {{Harry}}}\ and\ \bibinfo {author} {\bibnamefont {{LIGO Scientific
  Collaboration}}},\ }\href {\doibase 10.1088/0264-9381/27/8/084006} {\bibfield
   {journal} {\bibinfo  {journal} {Classical and Quantum Gravity}\ }\textbf
  {\bibinfo {volume} {27}},\ \bibinfo {eid} {084006} (\bibinfo {year}
  {2010})}\BibitemShut {NoStop}%
\bibitem [{\citenamefont {{LIGO Scientific Collaboration}}\ \emph
  {et~al.}(2013)\citenamefont {{LIGO Scientific Collaboration}}, \citenamefont
  {{Virgo Collaboration}}, \citenamefont {{Aasi}}, \citenamefont {{Abadie}},
  \citenamefont {{Abbott}}, \citenamefont {{Abbott}}, \citenamefont {{Abbott}},
  \citenamefont {{Abernathy}}, \citenamefont {{Accadia}}, \citenamefont
  {{Acernese}},\ and\ \citenamefont {et~al.}}]{2013arXiv1304.0670L}%
  \BibitemOpen
  \bibfield  {author} {\bibinfo {author} {\bibnamefont {{LIGO Scientific
  Collaboration}}}, \bibinfo {author} {\bibnamefont {{Virgo Collaboration}}},
  \bibinfo {author} {\bibfnamefont {J.}~\bibnamefont {{Aasi}}}, \bibinfo
  {author} {\bibfnamefont {J.}~\bibnamefont {{Abadie}}}, \bibinfo {author}
  {\bibfnamefont {B.~P.}\ \bibnamefont {{Abbott}}}, \bibinfo {author}
  {\bibfnamefont {R.}~\bibnamefont {{Abbott}}}, \bibinfo {author}
  {\bibfnamefont {T.~D.}\ \bibnamefont {{Abbott}}}, \bibinfo {author}
  {\bibfnamefont {M.}~\bibnamefont {{Abernathy}}}, \bibinfo {author}
  {\bibfnamefont {T.}~\bibnamefont {{Accadia}}}, \bibinfo {author}
  {\bibfnamefont {F.}~\bibnamefont {{Acernese}}}, \ and\ \bibinfo {author}
  {\bibnamefont {et~al.}},\ }\href@noop {} {\bibfield  {journal} {\bibinfo
  {journal} {ArXiv e-prints}\ } (\bibinfo {year} {2013})},\ \Eprint
  {http://arxiv.org/abs/1304.0670} {arXiv:1304.0670 [gr-qc]} \BibitemShut
  {NoStop}%
\bibitem [{\citenamefont {{Abadie}}\ \emph {et~al.}(2010)\citenamefont
  {{Abadie}}, \citenamefont {{Abbott}}, \citenamefont {{Abbott}}, \citenamefont
  {{Abernathy}}, \citenamefont {{Accadia}}, \citenamefont {{Acernese}},
  \citenamefont {{Adams}}, \citenamefont {{Adhikari}}, \citenamefont {{Ajith}},
  \citenamefont {{Allen}},\ and\ \citenamefont {et~al.}}]{2010CQGra..27q3001A}%
  \BibitemOpen
  \bibfield  {author} {\bibinfo {author} {\bibfnamefont {J.}~\bibnamefont
  {{Abadie}}}, \bibinfo {author} {\bibfnamefont {B.~P.}\ \bibnamefont
  {{Abbott}}}, \bibinfo {author} {\bibfnamefont {R.}~\bibnamefont {{Abbott}}},
  \bibinfo {author} {\bibfnamefont {M.}~\bibnamefont {{Abernathy}}}, \bibinfo
  {author} {\bibfnamefont {T.}~\bibnamefont {{Accadia}}}, \bibinfo {author}
  {\bibfnamefont {F.}~\bibnamefont {{Acernese}}}, \bibinfo {author}
  {\bibfnamefont {C.}~\bibnamefont {{Adams}}}, \bibinfo {author} {\bibfnamefont
  {R.}~\bibnamefont {{Adhikari}}}, \bibinfo {author} {\bibfnamefont
  {P.}~\bibnamefont {{Ajith}}}, \bibinfo {author} {\bibfnamefont
  {B.}~\bibnamefont {{Allen}}}, \ and\ \bibinfo {author} {\bibnamefont
  {et~al.}},\ }\href {\doibase 10.1088/0264-9381/27/17/173001} {\bibfield
  {journal} {\bibinfo  {journal} {Classical and Quantum Gravity}\ }\textbf
  {\bibinfo {volume} {27}},\ \bibinfo {eid} {173001} (\bibinfo {year}
  {2010})},\ \Eprint {http://arxiv.org/abs/1003.2480} {arXiv:1003.2480
  [astro-ph.HE]} \BibitemShut {NoStop}%
\bibitem [{\citenamefont {{Flanagan}}\ and\ \citenamefont
  {{Hughes}}(2005)}]{2005NJPh....7..204F}%
  \BibitemOpen
  \bibfield  {author} {\bibinfo {author} {\bibfnamefont {{\'E}.~{\'E}.}\
  \bibnamefont {{Flanagan}}}\ and\ \bibinfo {author} {\bibfnamefont {S.~A.}\
  \bibnamefont {{Hughes}}},\ }\href {\doibase 10.1088/1367-2630/7/1/204}
  {\bibfield  {journal} {\bibinfo  {journal} {New Journal of Physics}\ }\textbf
  {\bibinfo {volume} {7}},\ \bibinfo {pages} {204} (\bibinfo {year} {2005})},\
  \Eprint {http://arxiv.org/abs/gr-qc/0501041} {gr-qc/0501041} \BibitemShut
  {NoStop}%
\bibitem [{\citenamefont {{Carroll}}(2004)}]{BhagavadGita}%
  \BibitemOpen
  \bibfield  {author} {\bibinfo {author} {\bibfnamefont {S.~M.}\ \bibnamefont
  {{Carroll}}},\ }\href@noop {} {\emph {\bibinfo {title} {Spacetime and
  geometry / Sean Carroll.~San Francisco, CA, USA: Addison Wesley, ISBN
  0-8053-8732-3, 2004, XIV + 513 pp.}}}\ (\bibinfo {year} {2004})\BibitemShut
  {NoStop}%
\bibitem [{\citenamefont {{Adhikari}}(2014)}]{2014RvMP...86..121A}%
  \BibitemOpen
  \bibfield  {author} {\bibinfo {author} {\bibfnamefont {R.~X.}\ \bibnamefont
  {{Adhikari}}},\ }\href {\doibase 10.1103/RevModPhys.86.121} {\bibfield
  {journal} {\bibinfo  {journal} {Reviews of Modern Physics}\ }\textbf
  {\bibinfo {volume} {86}},\ \bibinfo {pages} {121} (\bibinfo {year} {2014})},\
  \Eprint {http://arxiv.org/abs/1305.5188} {arXiv:1305.5188 [gr-qc]}
  \BibitemShut {NoStop}%
\bibitem [{\citenamefont {Rubbo}\ \emph {et~al.}(2004)\citenamefont {Rubbo},
  \citenamefont {Cornish},\ and\ \citenamefont {Poujade}}]{rubbo2004}%
  \BibitemOpen
  \bibfield  {author} {\bibinfo {author} {\bibfnamefont {L.~J.}\ \bibnamefont
  {Rubbo}}, \bibinfo {author} {\bibfnamefont {N.~J.}\ \bibnamefont {Cornish}},
  \ and\ \bibinfo {author} {\bibfnamefont {O.}~\bibnamefont {Poujade}},\ }\href
  {\doibase 10.1103/PhysRevD.69.082003} {\bibfield  {journal} {\bibinfo
  {journal} {Phys. Rev. D}\ }\textbf {\bibinfo {volume} {69}},\ \bibinfo
  {pages} {082003} (\bibinfo {year} {2004})}\BibitemShut {NoStop}%
\bibitem [{\citenamefont {{Amaro-Seoane}}\ \emph {et~al.}(2013)\citenamefont
  {{Amaro-Seoane}}, \citenamefont {{Aoudia}}, \citenamefont {{Babak}},
  \citenamefont {{Bin{\'e}truy}}, \citenamefont {{Berti}}, \citenamefont
  {{Boh{\'e}}}, \citenamefont {{Caprini}}, \citenamefont {{Colpi}},
  \citenamefont {{Cornish}}, \citenamefont {{Danzmann}}, \citenamefont
  {{Dufaux}}, \citenamefont {{Gair}}, \citenamefont {{Hinder}}, \citenamefont
  {{Jennrich}}, \citenamefont {{Jetzer}}, \citenamefont {{Klein}},
  \citenamefont {{Lang}}, \citenamefont {{Lobo}}, \citenamefont {{Littenberg}},
  \citenamefont {{McWilliams}}, \citenamefont {{Nelemans}}, \citenamefont
  {{Petiteau}}, \citenamefont {{Porter}}, \citenamefont {{Schutz}},
  \citenamefont {{Sesana}}, \citenamefont {{Stebbins}}, \citenamefont
  {{Sumner}}, \citenamefont {{Vallisneri}}, \citenamefont {{Vitale}},
  \citenamefont {{Volonteri}}, \citenamefont {{Ward}},\ and\ \citenamefont
  {{Wardell}}}]{2013GWN.....6....4A}%
  \BibitemOpen
  \bibfield  {author} {\bibinfo {author} {\bibfnamefont {P.}~\bibnamefont
  {{Amaro-Seoane}}}, \bibinfo {author} {\bibfnamefont {S.}~\bibnamefont
  {{Aoudia}}}, \bibinfo {author} {\bibfnamefont {S.}~\bibnamefont {{Babak}}},
  \bibinfo {author} {\bibfnamefont {P.}~\bibnamefont {{Bin{\'e}truy}}},
  \bibinfo {author} {\bibfnamefont {E.}~\bibnamefont {{Berti}}}, \bibinfo
  {author} {\bibfnamefont {A.}~\bibnamefont {{Boh{\'e}}}}, \bibinfo {author}
  {\bibfnamefont {C.}~\bibnamefont {{Caprini}}}, \bibinfo {author}
  {\bibfnamefont {M.}~\bibnamefont {{Colpi}}}, \bibinfo {author} {\bibfnamefont
  {N.~J.}\ \bibnamefont {{Cornish}}}, \bibinfo {author} {\bibfnamefont
  {K.}~\bibnamefont {{Danzmann}}}, \bibinfo {author} {\bibfnamefont {J.-F.}\
  \bibnamefont {{Dufaux}}}, \bibinfo {author} {\bibfnamefont {J.}~\bibnamefont
  {{Gair}}}, \bibinfo {author} {\bibfnamefont {I.}~\bibnamefont {{Hinder}}},
  \bibinfo {author} {\bibfnamefont {O.}~\bibnamefont {{Jennrich}}}, \bibinfo
  {author} {\bibfnamefont {P.}~\bibnamefont {{Jetzer}}}, \bibinfo {author}
  {\bibfnamefont {A.}~\bibnamefont {{Klein}}}, \bibinfo {author} {\bibfnamefont
  {R.~N.}\ \bibnamefont {{Lang}}}, \bibinfo {author} {\bibfnamefont
  {A.}~\bibnamefont {{Lobo}}}, \bibinfo {author} {\bibfnamefont
  {T.}~\bibnamefont {{Littenberg}}}, \bibinfo {author} {\bibfnamefont {S.~T.}\
  \bibnamefont {{McWilliams}}}, \bibinfo {author} {\bibfnamefont
  {G.}~\bibnamefont {{Nelemans}}}, \bibinfo {author} {\bibfnamefont
  {A.}~\bibnamefont {{Petiteau}}}, \bibinfo {author} {\bibfnamefont {E.~K.}\
  \bibnamefont {{Porter}}}, \bibinfo {author} {\bibfnamefont {B.~F.}\
  \bibnamefont {{Schutz}}}, \bibinfo {author} {\bibfnamefont {A.}~\bibnamefont
  {{Sesana}}}, \bibinfo {author} {\bibfnamefont {R.}~\bibnamefont
  {{Stebbins}}}, \bibinfo {author} {\bibfnamefont {T.}~\bibnamefont
  {{Sumner}}}, \bibinfo {author} {\bibfnamefont {M.}~\bibnamefont
  {{Vallisneri}}}, \bibinfo {author} {\bibfnamefont {S.}~\bibnamefont
  {{Vitale}}}, \bibinfo {author} {\bibfnamefont {M.}~\bibnamefont
  {{Volonteri}}}, \bibinfo {author} {\bibfnamefont {H.}~\bibnamefont {{Ward}}},
  \ and\ \bibinfo {author} {\bibfnamefont {B.}~\bibnamefont {{Wardell}}},\
  }\href@noop {} {\bibfield  {journal} {\bibinfo  {journal} {GW Notes, Vol.~6,
  p.~4-110}\ }\textbf {\bibinfo {volume} {6}},\ \bibinfo {pages} {4} (\bibinfo
  {year} {2013})},\ \Eprint {http://arxiv.org/abs/1201.3621} {arXiv:1201.3621
  [astro-ph.CO]} \BibitemShut {NoStop}%
\bibitem [{\citenamefont {eLisa Consortium}(2013)}]{2013arXiv1305.5720C}%
  \BibitemOpen
  \bibfield  {author} {\bibinfo {author} {\bibnamefont {eLisa Consortium}},\
  }\href@noop {} {\bibfield  {journal} {\bibinfo  {journal} {ArXiv e-prints}\ }
  (\bibinfo {year} {2013})},\ \Eprint {http://arxiv.org/abs/1305.5720}
  {arXiv:1305.5720 [astro-ph.CO]} \BibitemShut {NoStop}%
\bibitem [{\citenamefont {eLISA consortium}(2013)}]{elisa}%
  \BibitemOpen
  \bibfield  {author} {\bibinfo {author} {\bibfnamefont {T.}~\bibnamefont
  {eLISA consortium}},\ }\href {https://www.elisascience.org/} {\enquote
  {\bibinfo {title} {Selected: The gravitational universe},}\ } (\bibinfo
  {year} {2013})\BibitemShut {NoStop}%
\bibitem [{\citenamefont {{Estabrook}}\ and\ \citenamefont
  {{Wahlquist}}(1975)}]{1975GReGr...6..439E}%
  \BibitemOpen
  \bibfield  {author} {\bibinfo {author} {\bibfnamefont {F.~B.}\ \bibnamefont
  {{Estabrook}}}\ and\ \bibinfo {author} {\bibfnamefont {H.~D.}\ \bibnamefont
  {{Wahlquist}}},\ }\href {\doibase 10.1007/BF00762449} {\bibfield  {journal}
  {\bibinfo  {journal} {General Relativity and Gravitation}\ }\textbf {\bibinfo
  {volume} {6}},\ \bibinfo {pages} {439} (\bibinfo {year} {1975})}\BibitemShut
  {NoStop}%
\bibitem [{\citenamefont {{Iorio}}(2013)}]{2013CQGra..30s5011I}%
  \BibitemOpen
  \bibfield  {author} {\bibinfo {author} {\bibfnamefont {L.}~\bibnamefont
  {{Iorio}}},\ }\href {\doibase 10.1088/0264-9381/30/19/195011} {\bibfield
  {journal} {\bibinfo  {journal} {Classical and Quantum Gravity}\ }\textbf
  {\bibinfo {volume} {30}},\ \bibinfo {eid} {195011} (\bibinfo {year}
  {2013})},\ \Eprint {http://arxiv.org/abs/1302.6920} {arXiv:1302.6920 [gr-qc]}
  \BibitemShut {NoStop}%
\bibitem [{\citenamefont {{Giorgetta, F. R.{\it et al.}}}(2013)}]{Giorgetta}%
  \BibitemOpen
  \bibfield  {author} {\bibinfo {author} {\bibnamefont {{Giorgetta, F. R.{\it
  et al.}}}},\ }\href {\doibase 10.1038/nphoton.2013.69} {\bibfield  {journal}
  {\bibinfo  {journal} {Nature Photonics}\ }\textbf {\bibinfo {volume} {7}},\
  \bibinfo {pages} {434} (\bibinfo {year} {2013})},\ \Eprint
  {http://arxiv.org/abs/1211.4902} {arXiv:1211.4902 [optics]} \BibitemShut
  {NoStop}%
\bibitem [{\citenamefont {Ang\'elil}\ \emph {et~al.}(2014)\citenamefont
  {Ang\'elil}, \citenamefont {Saha}, \citenamefont {Bondarescu}, \citenamefont
  {Jetzer}, \citenamefont {Sch\"arer},\ and\ \citenamefont
  {Lundgren}}]{earth_clox}%
  \BibitemOpen
  \bibfield  {author} {\bibinfo {author} {\bibfnamefont {R.}~\bibnamefont
  {Ang\'elil}}, \bibinfo {author} {\bibfnamefont {P.}~\bibnamefont {Saha}},
  \bibinfo {author} {\bibfnamefont {R.}~\bibnamefont {Bondarescu}}, \bibinfo
  {author} {\bibfnamefont {P.}~\bibnamefont {Jetzer}}, \bibinfo {author}
  {\bibfnamefont {A.}~\bibnamefont {Sch\"arer}}, \ and\ \bibinfo {author}
  {\bibfnamefont {A.}~\bibnamefont {Lundgren}},\ }\href {\doibase
  10.1103/PhysRevD.89.064067} {\bibfield  {journal} {\bibinfo  {journal} {Phys.
  Rev. D}\ }\textbf {\bibinfo {volume} {89}},\ \bibinfo {pages} {064067}
  (\bibinfo {year} {2014})}\BibitemShut {NoStop}%
\bibitem [{\citenamefont {Finn}\ and\ \citenamefont
  {Lommen}(2010)}]{FinnLommen2010}%
  \BibitemOpen
  \bibfield  {author} {\bibinfo {author} {\bibfnamefont {L.~S.}\ \bibnamefont
  {Finn}}\ and\ \bibinfo {author} {\bibfnamefont {A.~N.}\ \bibnamefont
  {Lommen}},\ }\href {http://stacks.iop.org/0004-637X/718/i=2/a=1400}
  {\bibfield  {journal} {\bibinfo  {journal} {The Astrophysical Journal}\
  }\textbf {\bibinfo {volume} {718}},\ \bibinfo {pages} {1400} (\bibinfo {year}
  {2010})}\BibitemShut {NoStop}%
\bibitem [{\citenamefont {{Detweiler}}(1979)}]{1979ApJ...234.1100D}%
  \BibitemOpen
  \bibfield  {author} {\bibinfo {author} {\bibfnamefont {S.}~\bibnamefont
  {{Detweiler}}},\ }\href {\doibase 10.1086/157593} {\bibfield  {journal}
  {\bibinfo  {journal} {\apj}\ }\textbf {\bibinfo {volume} {234}},\ \bibinfo
  {pages} {1100} (\bibinfo {year} {1979})}\BibitemShut {NoStop}%
\bibitem [{\citenamefont {{Lesovik}}\ \emph {et~al.}(2005)\citenamefont
  {{Lesovik}}, \citenamefont {{Lebedev}}, \citenamefont {{Mounutcharyan}},\
  and\ \citenamefont {{Martin}}}]{2005PhRvD..71l2001L}%
  \BibitemOpen
  \bibfield  {author} {\bibinfo {author} {\bibfnamefont {G.~B.}\ \bibnamefont
  {{Lesovik}}}, \bibinfo {author} {\bibfnamefont {A.~V.}\ \bibnamefont
  {{Lebedev}}}, \bibinfo {author} {\bibfnamefont {V.}~\bibnamefont
  {{Mounutcharyan}}}, \ and\ \bibinfo {author} {\bibfnamefont {T.}~\bibnamefont
  {{Martin}}},\ }\href {\doibase 10.1103/PhysRevD.71.122001} {\bibfield
  {journal} {\bibinfo  {journal} {\prd}\ }\textbf {\bibinfo {volume} {71}},\
  \bibinfo {eid} {122001} (\bibinfo {year} {2005})},\ \Eprint
  {http://arxiv.org/abs/astro-ph/0506602} {astro-ph/0506602} \BibitemShut
  {NoStop}%
\bibitem [{\citenamefont {{Cornish}}(2009)}]{2009PhRvD..80h7101C}%
  \BibitemOpen
  \bibfield  {author} {\bibinfo {author} {\bibfnamefont {N.~J.}\ \bibnamefont
  {{Cornish}}},\ }\href {\doibase 10.1103/PhysRevD.80.087101} {\bibfield
  {journal} {\bibinfo  {journal} {\prd}\ }\textbf {\bibinfo {volume} {80}},\
  \bibinfo {eid} {087101} (\bibinfo {year} {2009})},\ \Eprint
  {http://arxiv.org/abs/0910.4372} {arXiv:0910.4372 [gr-qc]} \BibitemShut
  {NoStop}%
\bibitem [{\citenamefont {{Rakhmanov}}(2009)}]{2009CQGra..26o5010R}%
  \BibitemOpen
  \bibfield  {author} {\bibinfo {author} {\bibfnamefont {M.}~\bibnamefont
  {{Rakhmanov}}},\ }\href {\doibase 10.1088/0264-9381/26/15/155010} {\bibfield
  {journal} {\bibinfo  {journal} {Classical and Quantum Gravity}\ }\textbf
  {\bibinfo {volume} {26}},\ \bibinfo {eid} {155010} (\bibinfo {year}
  {2009})}\BibitemShut {NoStop}%
\bibitem [{\citenamefont {{Finn}}(2009)}]{2009PhRvD..79b2002F}%
  \BibitemOpen
  \bibfield  {author} {\bibinfo {author} {\bibfnamefont {L.~S.}\ \bibnamefont
  {{Finn}}},\ }\href {\doibase 10.1103/PhysRevD.79.022002} {\bibfield
  {journal} {\bibinfo  {journal} {\prd}\ }\textbf {\bibinfo {volume} {79}},\
  \bibinfo {eid} {022002} (\bibinfo {year} {2009})},\ \Eprint
  {http://arxiv.org/abs/0810.4529} {arXiv:0810.4529 [gr-qc]} \BibitemShut
  {NoStop}%
\bibitem [{\citenamefont {{Koop}}\ and\ \citenamefont
  {{Finn}}(2014)}]{2014PhRvD..90f2002K}%
  \BibitemOpen
  \bibfield  {author} {\bibinfo {author} {\bibfnamefont {M.~J.}\ \bibnamefont
  {{Koop}}}\ and\ \bibinfo {author} {\bibfnamefont {L.~S.}\ \bibnamefont
  {{Finn}}},\ }\href {\doibase 10.1103/PhysRevD.90.062002} {\bibfield
  {journal} {\bibinfo  {journal} {\prd}\ }\textbf {\bibinfo {volume} {90}},\
  \bibinfo {eid} {062002} (\bibinfo {year} {2014})}\BibitemShut {NoStop}%
\bibitem [{\citenamefont {{Meers}}(1989)}]{1989PhLA..142..465M}%
  \BibitemOpen
  \bibfield  {author} {\bibinfo {author} {\bibfnamefont {B.~J.}\ \bibnamefont
  {{Meers}}},\ }\href {\doibase 10.1016/0375-9601(89)90515-X} {\bibfield
  {journal} {\bibinfo  {journal} {Physics Letters A}\ }\textbf {\bibinfo
  {volume} {142}},\ \bibinfo {pages} {465} (\bibinfo {year}
  {1989})}\BibitemShut {NoStop}%
\bibitem [{\citenamefont {{Lobo}}(1992)}]{1992CQGra...9.1385L}%
  \BibitemOpen
  \bibfield  {author} {\bibinfo {author} {\bibfnamefont {J.~A.}\ \bibnamefont
  {{Lobo}}},\ }\href {\doibase 10.1088/0264-9381/9/5/019} {\bibfield  {journal}
  {\bibinfo  {journal} {Classical and Quantum Gravity}\ }\textbf {\bibinfo
  {volume} {9}},\ \bibinfo {pages} {1385} (\bibinfo {year} {1992})}\BibitemShut
  {NoStop}%
\bibitem [{\citenamefont {{Cooperstock}}\ and\ \citenamefont
  {{Faraoni}}(1993)}]{1993CQGra..10.1189C}%
  \BibitemOpen
  \bibfield  {author} {\bibinfo {author} {\bibfnamefont {F.~I.}\ \bibnamefont
  {{Cooperstock}}}\ and\ \bibinfo {author} {\bibfnamefont {V.}~\bibnamefont
  {{Faraoni}}},\ }\href {\doibase 10.1088/0264-9381/10/6/016} {\bibfield
  {journal} {\bibinfo  {journal} {Classical and Quantum Gravity}\ }\textbf
  {\bibinfo {volume} {10}},\ \bibinfo {pages} {1189} (\bibinfo {year}
  {1993})},\ \Eprint {http://arxiv.org/abs/astro-ph/9303018} {astro-ph/9303018}
  \BibitemShut {NoStop}%
\bibitem [{\citenamefont {{Melissinos}}\ and\ \citenamefont
  {{Das}}(2010)}]{2010AmJPh..78.1160M}%
  \BibitemOpen
  \bibfield  {author} {\bibinfo {author} {\bibfnamefont {A.}~\bibnamefont
  {{Melissinos}}}\ and\ \bibinfo {author} {\bibfnamefont {A.}~\bibnamefont
  {{Das}}},\ }\href {\doibase 10.1119/1.3443566} {\bibfield  {journal}
  {\bibinfo  {journal} {American Journal of Physics}\ }\textbf {\bibinfo
  {volume} {78}},\ \bibinfo {pages} {1160} (\bibinfo {year} {2010})},\ \Eprint
  {http://arxiv.org/abs/1002.0809} {arXiv:1002.0809 [gr-qc]} \BibitemShut
  {NoStop}%
\bibitem [{\citenamefont {{Misner}}\ \emph {et~al.}(1973)\citenamefont
  {{Misner}}, \citenamefont {{Thorne}},\ and\ \citenamefont
  {{Wheeler}}}]{necronomicon}%
  \BibitemOpen
  \bibfield  {author} {\bibinfo {author} {\bibfnamefont {C.~W.}\ \bibnamefont
  {{Misner}}}, \bibinfo {author} {\bibfnamefont {K.~S.}\ \bibnamefont
  {{Thorne}}}, \ and\ \bibinfo {author} {\bibfnamefont {J.~A.}\ \bibnamefont
  {{Wheeler}}},\ }\href@noop {} {\emph {\bibinfo {title} {San Francisco:
  W.H.~Freeman and Co., 1973}}},\ edited by\ \bibinfo {editor} {\bibnamefont
  {{Misner, C.~W., Thorne, K.~S., \& Wheeler, J.~A.}}}\ (\bibinfo {year}
  {1973})\ p.\ \bibinfo {pages} {654}\BibitemShut {NoStop}%
\bibitem [{\citenamefont {{Maggiore}}(2008)}]{maggiore}%
  \BibitemOpen
  \bibfield  {author} {\bibinfo {author} {\bibfnamefont {M.}~\bibnamefont
  {{Maggiore}}},\ }\href@noop {} {\emph {\bibinfo {title} {{Gravitational
  Waves: Theory and Experiments: Volume 1.}}}},\ edited by\ \bibinfo {editor}
  {\bibnamefont {{Oxford University Press}}}\ (\bibinfo {year} {2008})\ p.\
  \bibinfo {pages} {495}\BibitemShut {NoStop}%
\bibitem [{\citenamefont {{Saulson}}(1997)}]{1997AmJPh..65..501S}%
  \BibitemOpen
  \bibfield  {author} {\bibinfo {author} {\bibfnamefont {P.~R.}\ \bibnamefont
  {{Saulson}}},\ }\href {\doibase 10.1119/1.18578} {\bibfield  {journal}
  {\bibinfo  {journal} {American Journal of Physics}\ }\textbf {\bibinfo
  {volume} {65}},\ \bibinfo {pages} {501} (\bibinfo {year} {1997})}\BibitemShut
  {NoStop}%
\bibitem [{\citenamefont {Faraoni}(2007)}]{faoroni}%
  \BibitemOpen
  \bibfield  {author} {\bibinfo {author} {\bibfnamefont {V.}~\bibnamefont
  {Faraoni}},\ }\href {\doibase 10.1007/s10714-007-0415-5} {\bibfield
  {journal} {\bibinfo  {journal} {General Relativity and Gravitation}\ }\textbf
  {\bibinfo {volume} {39}},\ \bibinfo {pages} {677} (\bibinfo {year}
  {2007})}\BibitemShut {NoStop}%
\bibitem [{\citenamefont {Alcubierre}(1994)}]{warpdrive}%
  \BibitemOpen
  \bibfield  {author} {\bibinfo {author} {\bibfnamefont {M.}~\bibnamefont
  {Alcubierre}},\ }\href {http://stacks.iop.org/0264-9381/11/i=5/a=001}
  {\bibfield  {journal} {\bibinfo  {journal} {Classical and Quantum Gravity}\
  }\textbf {\bibinfo {volume} {11}},\ \bibinfo {pages} {L73} (\bibinfo {year}
  {1994})}\BibitemShut {NoStop}%
\bibitem [{Note1()}]{Note1}%
  \BibitemOpen
  \bibinfo {note} {This metric solution to the field equations is unrealistic
  because it requires energy density distributions which are unphysical because
  they violate responsible energy conditions.}\BibitemShut {Stop}%
\bibitem [{\citenamefont {Hunter}(2007)}]{matplotlib}%
  \BibitemOpen
  \bibfield  {author} {\bibinfo {author} {\bibfnamefont {J.~D.}\ \bibnamefont
  {Hunter}},\ }\href@noop {} {\bibfield  {journal} {\bibinfo  {journal}
  {Computing In Science \& Engineering}\ }\textbf {\bibinfo {volume} {9}},\
  \bibinfo {pages} {90} (\bibinfo {year} {2007})}\BibitemShut {NoStop}%
\bibitem [{\citenamefont {Tinto}\ and\ \citenamefont
  {Dhurandhar}(2005)}]{lrr-2005-4}%
  \BibitemOpen
  \bibfield  {author} {\bibinfo {author} {\bibfnamefont {M.}~\bibnamefont
  {Tinto}}\ and\ \bibinfo {author} {\bibfnamefont {S.~V.}\ \bibnamefont
  {Dhurandhar}},\ }\href {\doibase 10.12942/lrr-2005-4} {\bibfield  {journal}
  {\bibinfo  {journal} {Living Reviews in Relativity}\ }\textbf {\bibinfo
  {volume} {8}} (\bibinfo {year} {2005}),\ 10.12942/lrr-2005-4}\BibitemShut
  {NoStop}%
\end{thebibliography}%

\end{document}